\documentclass[12pt]{revtex4}
\usepackage{amsfonts}

\begin{document}

\title{Classical spinning particle assuming a covariant Hamiltonian}
\author{A. B\'{e}rard}
\email{aberard001@noos.fr}
\affiliation{L.P.M.C., Institut\ de\ Physique, 
             1 blvd. Fran\c{c}ois Arago, 57070, Metz, France}
\author{J. Lages}
\email{lages@cfif.ist.utl.pt}
\affiliation{
Center for Physics of Fundamental Interactions,
Instituto Superior T\'ecnico,
Av. Rovisco Pais, 1049-001 Lisboa, Portugal}
\author{H.\ Mohrbach}
\email{mohrbach@cerbere.u-strasbg.fr}
\affiliation{Institut Charles Sadron, CNRS UPR 022,
             6 rue Boussingault, 67083 Strasbourg Cedex, France}

\begin{abstract}
We consider a classical spinning particle in the frame of the
relativistic physics by means of a covariant Hamiltonian and of a
generalization of Poisson brackets which take into account the 
gauge fields.
We obtain different equations of motion and evolution in this context and we
compare our results with those of Bargmann-Michel-Telegdi. An extension to
the case of a curved space and a link towards quantum theory
are given at the end of the paper.
\end{abstract}

\maketitle

\section{Introduction}

The notion of spin has played an important role in quantum field theory and
more generally in quantum mechanics, the classical limit is often studied
particularly in practical physics. Independently the classical approach of
this concept, solved partly in a brilliant way by Bargmann-Michel-Telegdi
\cite{BARGMANN} , is an old problem and by consequence the study of the spin
of a particle in the relativistic domain is carried out in many textbooks on
group theory or electrodynamics.

It is well known that it is E. Cartan which created in 1913 the spinor
theory that Pauli rediscovered in 1927 in the context of the quantum
mechanics. There are many manners to introduce the notion of classical spin,
the principal ones are
probably those using the spinorial representations of
the rotations group. Another one is to introduce the moving group $P(3)$ of
the translations and the rotations in a three dimensional euclidean space.
The rotation around the gravity center of a physical system introduces, if its
properties are position and orientation independent (case where 
$P(3)=T(3)\times SO(3)$), the following relation for the total angular
momentum
\begin{equation}
\overrightarrow{J}=\overrightarrow{x}\wedge \overrightarrow{p}+
\overrightarrow{S}=\overrightarrow{L}+\overrightarrow{S},
\end{equation}
where $\overrightarrow{x}$ is the gravity center position, 
$\overrightarrow{L}$ is the vectorial orbital momentum and
$\overrightarrow{S}$ is the
vectorial intrinsic momentum which is usually connected with the classical
spin.

A similar relation is introduced in
the case of the restrained relativity for the
Poincar\'{e} group, the total tensorial angular momentum is then
\begin{equation}
J^{\mu \nu }=L^{\mu \nu }+S^{\mu \nu },
\end{equation}
where we have defined the tensorial orbital momentum by $L^{\mu \nu
}=x^{\mu }p^{\nu }-x^{\nu }p^{\mu }$. The tensor $S^{\mu \nu }$ being
connected to the quadri-dimensional spin.

Using the Noether's theorem which tell us that every continuous
transformation leaving invariant the action
associated to a physical system introduces a conserved quantity. In the case
where the infinitesimal transformations of the coordinates are the homogeneous
Lorentz transformations, the conserved quantity is then the total
tensorial angular momentum defined as in the group theory.

In section II we briefly describe the
formalism \cite{NOUS1,NOUS2,NOUS3} we will use to study the classical
spin. In particular we
characterize
the dynamical evolution by the means of a covariant Hamiltonian,
in a way very similar to that followed by Goldstein in the case of the
electromagnetism \cite{GOLDSTEIN} and of a
generalization of the Poisson brackets which takes fully into account
the gauge fields in its construction.
We will also insist on the fact that we will not work as usual
with the proper time but with another time-like parameter.
In section III, we use the formalism introduced before 
to construct the simplest spinor algebra which will enable us
to obtain in section IV the spin equations of motion. The two last sections
are devoted to a short extension of our formalism 
to the case of a curved space (section V) and to a link with
the quantum theory (section VI).
Before going further we mention that our generalized brackets are closely
connected to those
introduced by Feynman in his remarkable demonstration of Maxwell equations
where he tried to develop a quantization procedure without resort to a
Lagrangian or a Hamiltonian \cite{DYSON}.

\section{Basic formalism}

We introduce a non commutative algebra on a $M_{4}$ Minkowski space
by the mean of skew symmetric brackets defined by the following laws
\begin{equation}
\lbrack x^{\mu },y^{\nu }]=-[y^{\nu },x^{\mu }],
\end{equation}
\begin{equation}
\lbrack x^{\mu },y^{\nu }z^{\rho }]=[x^{\mu },y^{\nu }]z^{\rho }+[x^{\mu
},z^{\rho }]y^{\nu }\mbox{\,\,\, and \,\,\,\,}
[x^{\mu }y^{\nu },z^{\rho }]=[y^{\nu
},z^{\rho }]x^{\mu }+[x^{\mu },z^{\rho }]y^{\nu },
\end{equation}
and we require
also the locality property (case of the commutative geometry)
\begin{equation}
\left[ x^{\mu },x^{\nu }\right] =0.
\end{equation}
Let us define a one $\tau $ parameter group of diffeomorphisms 
\begin{equation}
g(\Bbb{R}\times M_{4}%
)\longrightarrow M_{4}:g(\tau ,x^{\mu })=g^{\tau }x^{\mu }=x^{\mu
}(\tau ),
\end{equation}
then the ''velocity vector'' associated to the particle is naturally
introduced by the dynamical equation
\begin{equation}
\stackrel{.}{x}^{\mu }=\frac{d}{d\tau }g^{\tau }x^{\mu },
\end{equation}
and defined by
\begin{equation}
\stackrel{.}{x}^{\mu }=\frac{dx^{\mu }}{d\tau }=\left[ x^{\mu },H\right]
\label{hamilton}
\end{equation}
where $H$ (the covariant Hamiltonian of this dynamics) is a priori an
expandable function of $x^{\mu }$ and $\stackrel{.}{x}^{\mu }$. 
We choose the Hamiltonian $H$ as the most simple covariant one \cite{NOUS1}
\begin{equation}
H=\frac{1}{2}mg_{\mu \nu }(x,\stackrel{.}{x})\stackrel{.}{x}^{\mu }\stackrel{%
.}{x}^{\nu },
\end{equation}
where $g^{\mu \nu }(x,\stackrel{.}{x})$ is the metric tensor of the
physical space which is chosen for the moment as a Minkowski flat space
\begin{equation}
g^{\mu \nu }(x,\stackrel{.}{x})\rightarrow g^{\mu \nu }=\left( 
\begin{array}{llll}
1 & 0 & 0 & 0 \\ 
0 & 1 & 0 & 0 \\ 
0 & 0 & 1 & 0 \\ 
0 & 0 & 0 & -1
\end{array}
\right).
\end{equation}
Equation (\ref{hamilton}) gives directly
\begin{equation}
\stackrel{.}{x}^{\mu }=\left[ x^{\mu },H\right] =m\left[ x^{\mu },\stackrel{.%
}{x}^{\nu }\right] \stackrel{.}{x}_{\nu },
\end{equation}
from where we deduce easily
\begin{equation}
\left[ x^{\mu },\stackrel{.}{x}^{\nu }\right] =\frac{1}{m}g^{\mu \nu }.
\label{métrique}
\end{equation}
Our construction shows clearly that this Hamiltonian is covariant in similar 
way to the four dimensional covariant
Hamiltonian introduced by Goldstein for electromagnetism \cite{GOLDSTEIN}.

We directly obtain for expandable functions in $x$ or $\stackrel{.}{x}$ the
following useful relations
\begin{equation}
\left\{ 
\begin{array}{c}
\left[ x^{\mu },f(x,\stackrel{.}{x})\right] =\left\{ x^{\mu },f(x,\stackrel{.%
}{x})\right\} =\frac{1}{m}\frac{\partial f(x,\stackrel{.}{x})}{\partial 
\stackrel{.}{x}_{\mu }}, \\ 
\\ 
\left[ \stackrel{.}{x}^{\mu },f(x,\stackrel{.}{x})\right] =\left\{ \stackrel{%
.}{x}^{\mu },f(x,\stackrel{.}{x})\right\} +\left[ \stackrel{.}{x}^{\mu },%
\stackrel{.}{x}^{\nu }\right] \frac{\partial f(x,\stackrel{.}{x})}{\partial 
\stackrel{.}{x}^{\nu }}=-\frac{1}{m}\frac{\partial f(x,\stackrel{.}{x})}{%
\partial x_{\mu }}+\left[ \stackrel{.}{x}^{\mu },\stackrel{.}{x}^{\nu
}\right] \frac{\partial f(x,\stackrel{.}{x})}{\partial \stackrel{.}{x}^{\nu }%
}, \\ 
\\ 
\left[ f(x,\stackrel{.}{x}),h(x,\stackrel{.}{x})\right] =\left\{ f(x,%
\stackrel{.}{x}),h(x,\stackrel{.}{x})\right\} +\left[ \stackrel{.}{x}^{\mu },%
\stackrel{.}{x}^{\nu }\right] \frac{\partial f(x,\stackrel{.}{x})}{\partial 
\stackrel{.}{x}^{\mu }}\frac{\partial h(x,\stackrel{.}{x})}{\partial 
\stackrel{.}{x}^{\nu }}.
\end{array}
\right.
\end{equation}
We use the usual Poisson brackets on functions defined
here on the tangent bundle space
\begin{equation}
\left\{ f(x,\stackrel{.}{x}),h(x,\stackrel{.}{x})\right\} =\frac{g^{\mu \nu }%
}{m}\left( \frac{\partial f(x,\stackrel{.}{x})}{\partial x^{\mu }}\frac{%
\partial h(x,\stackrel{.}{x})}{\partial \stackrel{.}{x}^{\nu }}-\frac{%
\partial f(x,\stackrel{.}{x})}{\partial \stackrel{.}{x}^{\nu }}\frac{%
\partial h(x,\stackrel{.}{x})}{\partial x^{\mu }}\right) .
\end{equation}

It is easy to check that, for a particle with an electric charge $q$, the
tensor $\left[ \stackrel{.}{x}^{\mu },\stackrel{.}{x}^{\nu }\right]$ is a
skew symmetric tensor and  we denote it as $\frac{q}{m^{2}}F^{\mu \nu }(x)$.
It corresponds to a generalization of the three dimensional electromagnetic
tensor introduced in a preceding paper \cite{NOUS1} .

A very important remark is, as in the Feynman approach, that the time
parameter is not the proper time but the conjugate coordinate of this
covariant Hamiltonian. To see this we borrow Tanimura's argument 
\cite{TANIMURA} which considers the relation 
\begin{equation}
\stackrel{.}{x}^{2}=g^{\mu \nu }\frac{dx^{\mu }}{dt_{p}}\frac{dx^{\nu }}{%
dt_{p}}=1,
\end{equation}
\smallskip which implies 
\begin{equation}
\left[ x^{\lambda },g^{\mu \nu }\frac{dx^{\mu }}{dt_{p}}\frac{dx^{\nu }}{%
dt_{p}}\right] =0,
\end{equation}
and is in contradiction with
\begin{equation}
\left[ x^{\lambda },g^{\mu \nu }\frac{dx^{\mu }}{d\tau }\frac{dx^{\nu }}{%
d\tau }\right] =\frac{2}{m}\stackrel{.}{x}^{\lambda }.
\end{equation}

In our case we obtain
\begin{eqnarray}
\stackrel{.}{x}^{2} &=&\left( \frac{dx^{0}}{d\tau }\right) ^{2}-\left( \frac{%
dx^{i}}{d\tau }\right) ^{2}=\left\{ \left( \frac{dx^{0}}{dt_{p}}\right)
^{2}-\left( \frac{dx^{i}}{dt_{p}}\right) ^{2}\right\} \left( \frac{dt_{p}}{%
d\tau }\right) ^{2}  \nonumber \\
&=&\left( \frac{dt_{p}}{d\tau }\right) ^{2}=\left( 1-\beta ^{2}\right)
\left( \frac{dt}{d\tau }\right) ^{2},
\end{eqnarray}
with the usual parameter
$\beta =\frac{v}{c}$. Another manner to introduce this
concept would be to follow for example Gill {\it et al.}
\cite{GILL1,GILL2} which
give in a rather close context
\begin{equation}
\left( \frac{dx}{d\tau }\right) ^{2}=\stackrel{.}{x}^{2}=\frac{2H}{m}.
\end{equation}

\section{Algebra structure with spinning particle}

Let the total angular momentum be 
\begin{equation}
J^{\mu \nu }\mathcal{=}L^{\mu \nu }\mathcal{+S}^{\mu \nu },
\end{equation}
where $L^{\mu \nu }$ is the angular momentum defined in the usual manner 
\begin{equation}
L^{\mu \nu }=m\left( x^{\mu }\dot{x}^{\nu }-x^{\nu }\dot{x}^{\mu }\right) ,
\end{equation}
and $\mathcal{S}^{\mu \nu }$ the spinning angular momentum. Naturally the
latter commutes with the external quantities 
\begin{equation}
\left\{ 
\begin{array}{c}
\left[ x^{\mu },\mathcal{S}^{\rho \sigma }\right] =0, \\ 
\\ 
\left[ \dot{x}^{\mu },\mathcal{S^{\rho \sigma }}\right] =0, \\ 
\\ 
\left[ L^{\mu \nu },\mathcal{S^{\rho \sigma }}\right] =0,
\end{array}
\right.
\end{equation}
and has to obey to the same algebra commutation relation as the generalized
angular momentum $\mathcal{L^{\mu \nu }}$ 
\begin{equation}
\left[ \mathcal{S}^{\mu \nu },\mathcal{S}^{\rho \sigma }\right] =g^{\nu \rho
}\mathcal{S}^{\mu \sigma }-g^{\nu \sigma }\mathcal{S}^{\mu \rho }+g^{\mu
\sigma }\mathcal{S}^{\nu \rho }-g^{\mu \rho }\mathcal{S}^{\nu \sigma }.
\end{equation}

\subsection{Definition of the spin 1-form}

We introduce the following differential 3-forms defined on a
quadri-dimensional Minkowski flat space 
\begin{equation}
W=3!\,m\,\dot{x}\wedge J,
\end{equation}
which becomes if we project it on a basis 
\begin{equation}
W^{\mu \nu \rho }=3\,m\,\dot{x}^{\mu }\Bbb{\,}\mathcal{S}^{\nu \rho },
\end{equation}
where the components of the spinning tensor are skew symmetrized.

The well known Pauli-Lubanski polarization 1-form is connected to this spin
1-form by the relation 
\begin{equation}
^{\ast }W=\,^{*}3!\left( m\,\dot{x}\wedge J\right) ,
\end{equation}
or using the definition of the Hodge duality on the components 
\begin{equation}
^{\ast }W^{\mu }=\frac{1}{3!}\varepsilon ^{\mu \nu \rho \sigma }W_{\nu \rho
\sigma }=\frac{m}{2}\varepsilon ^{\mu \nu \rho \sigma }\dot{x}_{\nu }%
\mathcal{S}_{\rho \sigma }=m\dot{x}_{\nu }{}^{*}\mathcal{S}^{\mu \nu }=m\Bbb{%
S}^{\mu }.  \label{Pauli}
\end{equation}
We naturally observe that in every frame $\Bbb{S}^{\mu }\dot{x}_{\mu }=0$, and
we deduce from this relation that the quadrivector $\Bbb{S}$ can be written in
the rest frame under the form $\Bbb{S}^{\mu }=\left( 0,\overrightarrow{\xi }%
\right)$. We have in every frame $\Bbb{S}^{2}=-\xi ^{2}$ , where $%
\overrightarrow{\xi }$ is closely related to the particle polarization.

As the time parameter is not the proper time, we can not use the relation 
\begin{equation}
\stackrel{.}{x}^{2}=1,
\end{equation}
\smallskip and have not the usual inverse relation of (\ref{Pauli})
\begin{equation}
\mathcal{S}^{\mu \nu }=\varepsilon ^{\mu \nu \rho \sigma }\stackrel{.}{x}
_{\rho }\Bbb{S}_{\sigma },
\end{equation}
but instead
\begin{equation}
\dot{x}^{\nu }\mathcal{S}^{\rho \sigma }+\dot{x}^{\rho }\mathcal{S}^{\sigma
\nu }+\dot{x}^{\sigma }\mathcal{S}^{\nu \rho }=\varepsilon ^{\mu \nu \rho
\sigma }\Bbb{S}_{\mu },  \label{inverse}
\end{equation}
where we use the formula
\begin{equation}
\varepsilon ^{\mu \nu \rho \sigma }\varepsilon ^{\alpha \beta \gamma \delta
}=-\det \left( g^{ab}\right) ,
\end{equation}
with $a=\mu \nu \rho \sigma $ , and: $b=\alpha \beta \gamma \delta $ .

\subsection{External and Intrinsic algebraic commutation relations}

We can derive the algebraic commutation relations between the spin 1-form $%
S^{\mu }$ and the external quantities

\begin{equation}
\left\{ 
\begin{array}{c}
\left[ \Bbb{S}^{\mu },x^{\nu }\right] =-\frac{1}{m}\left( ^{*}\mathcal{S}%
\right) ^{\mu \nu }, \\ 
\\ 
\left[ \Bbb{S}^{\mu },\dot{x}^{\nu }\right] =\frac{q}{m^{2}}\left( ^{*}%
\mathcal{S}F\right) ^{\mu \nu }, \\ 
\\ 
\left[ \Bbb{S}^{\mu },L^{\rho \sigma }\right] =\,^{*}\mathcal{S}^{\mu \sigma
}\dot{x}^{\rho }-\,^{*}\mathcal{S}^{\mu \rho }\dot{x}^{\sigma }+\frac{q}{m}%
\left\{ \left( ^{*}\mathcal{S}F\right) ^{\mu \sigma }x^{\rho }-\left( ^{*}%
\mathcal{S}F\right) ^{\mu \rho }x^{\sigma }\right\} ,
\end{array}
\right.
\end{equation}
and between the spin 1-form $\Bbb{S}^{\mu }$ and the intrinsic quantity $%
\mathcal{S}^{\mu \nu }$ 
\begin{equation}
\left[ \Bbb{S}^{\mu },\mathcal{S}^{\rho \sigma }\right] =g^{\mu \sigma }\Bbb{%
S}^{\rho }-g^{\mu \rho }\Bbb{S}^{\sigma }-\left( ^{*}\mathcal{S}^{\mu \rho }%
\stackrel{.}{x}^{\sigma }-^{*}\mathcal{S}^{\mu \sigma }\stackrel{.}{x}^{\rho
}\right) .  \label{[S,S]}
\end{equation}

Finally we obtain the commutation relation involving different components
of the spin 1-form $\Bbb{S}^{\mu }$ 
\begin{equation}
\left[ \Bbb{S}^{\mu },\Bbb{S}^{\nu }\right] =\varepsilon ^{\mu \nu
}{}_{\alpha \beta }\dot{x}^{\alpha }\Bbb{S}^{\beta }-\frac{q}{m^{2}}\left(
\,^{*}\mathcal{S}F\,^{*}\mathcal{S}\right) ^{\mu \nu }.
\end{equation}

\subsection{Poincar\'{e} momentum}

Let the total angular momentum be 
\begin{equation}
\mathcal{J^{\mu \nu }=L^{\mu \nu }+S}^{\mu \nu }
\end{equation}
where on one side $\mathcal{L^{\mu \nu }}=L^{\mu \nu }+M^{\mu \nu }$ is the
generalized angular momentum with the generalization of the Poincar\'{e}
\cite{POINCARE} magnetic angular momentum $M^{\mu \nu }$ defined as in our
preceding paper \cite{NOUS3} , and on the other $\mathcal{S}^{\mu \nu }$ is
the spinning angular momentum. Naturally this latter commutes with the
external quantities
\[
\left[ M^{\mu \nu },\mathcal{S^{\rho \sigma }}\right] =0.
\]

We recall the relation between the Poincar\'{e} momentum and the gauge field
\begin{equation}
g^{\mu \rho }M^{\nu \sigma }-g^{\nu \rho }M^{\mu \sigma }+g^{\mu \sigma
}M^{\rho \nu }-g^{\nu \sigma }M^{\rho \mu }=q(F^{\nu \sigma }x^{\mu }x^{\rho
}-F^{\mu \sigma }x^{\upsilon }x^{\rho }+F^{\rho \mu }x^{\mu }x^{\sigma
}-F^{\rho \mu }x^{\nu }x^{\sigma }),
\end{equation}
with the algebra equations
\begin{equation}
\left\{ 
\begin{array}{c}
\left[ x^{\mu },M^{\rho \sigma }\right] =0, \\ 
\\ 
\left[ \dot{x}^{\mu },M^{\rho \sigma }\right] =\frac{q}{m}\left( x^{\rho
}F^{\mu \sigma }-x^{\sigma }F^{\mu \rho }\right) , \\ 
\\ 
\left[ L^{\mu \nu },M^{\rho \sigma }\right] =\frac{q}{m}\left( F^{\nu \sigma
}x^{\mu }x^{\rho }-F^{\mu \sigma }x^{\upsilon }x^{\rho }+F^{\rho \mu }x^{\mu
}x^{\sigma }-F^{\rho \mu }x^{\nu }x^{\sigma }\right) \\ 
=g^{\mu \rho }M^{\nu \sigma }-g^{\nu \rho }M^{\mu \sigma }+g^{\mu \sigma
}M^{\rho \nu }-g^{\nu \sigma }M^{\rho \mu }, \\ 
\\ 
\left[ M^{\mu \nu },M^{\rho \sigma }\right] =0.
\end{array}
\right.
\end{equation}
Then we find
\begin{equation}
\left\{ 
\begin{array}{c}
\left[ \Bbb{S}^{\mu },M^{\rho \sigma }\right] =\frac{q}{m}\left\{ \left( ^{*}%
\mathcal{S}F\right) ^{\mu \rho }x^{\sigma }-\left( ^{*}\mathcal{S}F\right)
^{\mu \sigma }x^{\rho }\right\} , \\ 
\\ 
\left[ \Bbb{S}^{\mu },\mathcal{L}^{\rho \sigma }\right] =\,^{*}\mathcal{S}%
^{\mu \sigma }\dot{x}^{\rho }-\,^{*}\mathcal{S}^{\mu \rho }\dot{x}^{\sigma },
\end{array}
\right.
\end{equation}
and we are able to
write the commutation law between the spin 1-form $\Bbb{S}^{\mu }$
and the total angular momentum $\mathcal{J^{\nu \rho }}$ 
\begin{equation}
\left[ \Bbb{S}^{\mu },\mathcal{J}^{\rho \sigma }\right] =g^{\mu \sigma }\Bbb{%
S}^{\rho }-g^{\mu \rho }\Bbb{S}^{\sigma }.
\end{equation}



\section{Motion equations}

It is well known that the Maxwel equations and the Lorentz force law are not
sufficient to completely describe the behavior of a spinning particle and
that it is necessary to specify also an equation of motion for the spin
degrees of freedom.

\subsection{Simplest motion equation}

In Minkowski space, the simplest covariant Hamiltonian with a 
spin-electromagnetic field interaction term is
\begin{equation}
H=\frac{1}{2}m\stackrel{.}{x}^{2}-\frac{q}{2m}F\mathcal{S}=\frac{1}{2}m%
\stackrel{.}{x}^{\mu }\stackrel{.}{x}_{\mu }-\frac{q}{2m}F^{\mu \nu }%
\mathcal{S}_{\mu \nu }
\end{equation}
where the relation (\ref{métrique}) is still valid. We have
then the motion equations of a spinning classical particle
\begin{equation}
\stackrel{..}{x}^{\mu }=\left[ \stackrel{.}{x}^{\mu },H\right] =\frac{q}{m}%
F^{\mu \nu }\stackrel{.}{x}_{\nu }+\frac{q}{2m^{2}}\mathcal{S}_{\rho \sigma
}\partial ^{\mu }F^{\rho \sigma },
\end{equation}
of the spin tensor $S^{\mu \nu }$
\begin{equation}
\stackrel{.}{\mathcal{S}}^{\mu \nu }=\left[ \mathcal{S}^{\mu \nu },H\right] =%
\frac{q}{2m}F_{\rho \sigma }\left[ \mathcal{S}^{\rho \sigma },\mathcal{S}%
^{\mu \nu }\right] =\frac{q}{m}\left( F^{\mu }{}_{\rho }\mathcal{S}^{\rho
\nu }-\mathcal{S}^{\mu }{}_{\rho }F^{\rho \nu }\right) =\frac{q}{m}\left( F%
\mathcal{S}-\mathcal{S}F\right) ^{\mu \nu },
\end{equation}
and of the quadri-vector spin $\mathcal{S}^{\mu }$
\begin{eqnarray}
\stackrel{.}{\Bbb{S}}^{\mu } &=&\frac{1}{2}\varepsilon ^{\mu }{}_{\nu \rho
\sigma }\left[ \stackrel{.}{x}^{\nu }\mathcal{S}^{\rho \sigma
},H\right]  \nonumber \\
&=&\frac{q}{2m}\varepsilon ^{\mu \nu \rho \sigma }F_{\nu \alpha }%
\stackrel{.}{x}^{\alpha }\mathcal{S}_{\rho \sigma }+\frac{q}{4m^{2}}%
\varepsilon ^{\mu \nu \rho \sigma }\mathcal{S}_{\rho \sigma }%
\mathcal{S}^{\alpha \beta }\partial _{\nu }F_{\alpha \beta }  \nonumber \\
&&+\frac{q}{4m}\varepsilon ^{\mu }{}_{\nu \rho \sigma }\stackrel{.}{x%
}^{\nu }\left[ \mathcal{S}^{\rho \sigma },\mathcal{S}^{\alpha \beta }\right]
F_{\alpha \beta }.
\end{eqnarray}

We will now turn to the last term of the left side of this equation

\begin{equation}
\frac{q}{4m}\varepsilon ^{\mu }{}_{\nu \rho \sigma }\stackrel{.}{x}%
^{\nu }\left[ \mathcal{S}^{\rho \sigma },\mathcal{S}^{\alpha \beta }\right]
F_{\alpha \beta }=\frac{q}{2m}\varepsilon ^{\mu }{}_{\nu \rho \sigma }
\stackrel{.}{x}^{\nu }\left( F_{\alpha }{}^{\sigma }\mathcal{S}^{\alpha
\rho }-F_{\alpha }{}^{\rho }\mathcal{S}^{\alpha \sigma }\right) ,
\end{equation}
and then

\begin{eqnarray}
\stackrel{.}{\Bbb{S}}^{\mu } &=&\frac{q}{2m}\varepsilon ^{\mu \nu \rho
\sigma }F_{\nu \alpha }\stackrel{.}{x}^{\alpha }\mathcal{S}_{\rho
\sigma }+\frac{q}{2m}\varepsilon ^{\mu }{}_{\nu \rho \sigma }%
\stackrel{.}{x}^{\nu }\left( F_{\alpha }{}^{\sigma }\mathcal{S}^{\alpha \rho
}-F_{\alpha }{}^{\rho }\mathcal{S}^{\alpha \sigma }\right)  \nonumber \\
&&+\frac{q}{4m^{2}}\varepsilon ^{\mu \nu \rho \sigma }\mathcal{S}%
_{\rho \sigma }\mathcal{S}^{\alpha \beta }\partial _{\nu }F_{\alpha \beta } 
\nonumber \\
&=&\frac{q}{2m}F^{\mu \nu }\Bbb{S}_{\nu }+\frac{q}{4m^{2}}\varepsilon ^{\mu
\nu \rho \sigma }\mathcal{S}_{\rho \sigma }\mathcal{S}^{\alpha \beta
}\partial _{\nu }F_{\alpha \beta },
\end{eqnarray}
which is the equation found in another context by Van Holten\cite{VAN HOLTEN}
.

\subsection{Bargmann-Michel-Telegdi equation}

This classical relativistic equation which describes the behavior of the
spin quadri-vector can be generalized if we only require , like
Bargmann-Michel-Telegdi, the relativistic invariance.
This
generalized covariant Hamiltonian with interesting spin interaction terms
and linear terms in the field can be written {\it a priori} 
in the following form
\begin{equation}
H=\frac{1}{2}m\stackrel{.}{x}^{2}-\frac{q}{2m}F^{\rho \sigma }\mathcal{S}%
_{\rho \sigma }+a\stackrel{.}{x}_{\rho }F^{\rho \sigma }\mathcal{S}_{\sigma
\tau }\stackrel{.}{x}^{\tau }+....,
\end{equation}
which gives for the motion equation using
(\ref{métrique}),
\begin{equation}
\stackrel{..}{x}^{\mu }=\frac{q}{m}F^{\mu \nu }\stackrel{.}{x}_{\nu }+\frac{q%
}{2m^{2}}\mathcal{S}_{\rho \sigma }\partial ^{\mu }F^{\rho \sigma }-\frac{a}{%
m}\stackrel{.}{x}_{\rho }\partial ^{\mu }F^{\rho \sigma }\mathcal{S}_{\sigma
\tau }\stackrel{.}{x}^{\tau }+....
\end{equation}

The evolution equation of the spin tensor is then naturally modified
\begin{equation}
\stackrel{.}{\mathcal{S}}^{\mu \nu }=\frac{q}{m}\left( F^{\mu }{}_{\rho }%
\mathcal{S}^{\rho \nu }-\mathcal{S}^{\mu }{}_{\rho }F^{\rho \nu }\right)
+2a\left( \stackrel{.}{x}^{\mu }F^{\rho \nu }{}-\stackrel{.}{x}^{\nu
}F^{\rho \mu }\right) \stackrel{.}{x}_{\rho }.
\end{equation}
The evolution equation of the quadri-vector spin is 
\begin{equation}
\stackrel{.}{\Bbb{S}}^{\mu }=\frac{q}{m}F^{\mu \nu }\Bbb{S}_{\nu }+\frac{q}{%
4m^{2}}\varepsilon ^{\mu \nu \rho \sigma }\mathcal{S}_{\rho \sigma }%
\mathcal{S}^{\alpha \beta }\partial _{\nu }F_{\alpha \beta }+a\left[ 
\stackrel{.}{x}_{\rho }F^{\rho \sigma }\mathcal{S}_{\sigma \tau }\stackrel{.%
}{x}^{\tau },\Bbb{S}^{\mu }\right] +.....
\end{equation}
Keeping only linear terms the last term is transformed into
\begin{equation}
\left[ \Bbb{S}^{\mu },\stackrel{.}{x}_{\rho }F^{\rho \sigma }\mathcal{S}%
_{\sigma \tau }\stackrel{.}{x}^{\tau }\right] =-\frac{1}{2}\varepsilon ^{\mu
\alpha \beta \gamma }\left\{ \frac{1}{m}\mathcal{S}_{\beta \gamma }\stackrel{%
.}{x}_{\rho }\partial _{\alpha }F^{\rho \sigma }\mathcal{S}_{\sigma \tau }%
\stackrel{.}{x}^{\tau }+\stackrel{.}{x}_{\alpha }\left[ \mathcal{S}_{\beta
\gamma },\mathcal{S}_{\sigma \tau }\right] \stackrel{.}{x}_{\rho }F^{\rho
\sigma }\stackrel{.}{x}^{\tau }\right\} ,
\end{equation}
where

\begin{equation}
\frac{1}{2}\varepsilon ^{\mu \alpha \beta \gamma }\stackrel{.}{x}_{\alpha
}\left[ \mathcal{S}_{\beta \gamma },\mathcal{S}_{\sigma \tau }\right] 
\stackrel{.}{x}_{\rho }F^{\rho \sigma }\stackrel{.}{x}^{\tau }=-\varepsilon
^{\mu \alpha \beta \gamma }\stackrel{.}{x}^{2}\stackrel{.}{x}_{\alpha
}F^{\rho }{}_{\gamma }\mathcal{S}_{\beta \rho }-2\stackrel{.}{x}^{2}F^{\mu
\nu }{}\Bbb{S}_{\nu }-2\left( \stackrel{.}{x}F\Bbb{S}\right) \stackrel{.}{x}%
^{\mu }.
\end{equation}

Finally the evolution equation is
\begin{eqnarray}
\stackrel{.}{\Bbb{S}}^{\mu } &=&\left( \frac{q}{m}+2a\stackrel{.}{x}%
^{2}\right) F^{\mu \nu }\Bbb{S}_{\nu }-2a\left( \Bbb{S}F\stackrel{.}{x}%
\right) \stackrel{.}{x}^{\mu }+\frac{q}{2m^{2}}\left( ^{*}\mathcal{S}%
\right) ^{\mu \nu }\mathcal{S}^{\alpha \beta }\partial _{\nu }F_{\alpha
\beta }  \nonumber \\
&&-\frac{a}{2}\left\{ \frac{1}{m}\left( ^{*}\mathcal{S}\right) ^{\mu \alpha
}\partial _{\alpha }F^{\rho \sigma }\mathcal{S}_{\sigma \tau }\stackrel{.}{x}%
^{\tau }\stackrel{.}{x}_{\rho }-4^{*}\left( F\mathcal{S}\right) ^{\mu \alpha
}\stackrel{.}{x}_{\alpha }\stackrel{.}{x}^{2}\right\}.
\end{eqnarray}
In the limit that we can rewrite the above equation for weak constant fields
\begin{eqnarray}
\stackrel{.}{\Bbb{S}}^{\mu } &=&\left( \frac{q}{m}+2a\stackrel{.}{x}%
^{2}\right) F^{\mu \nu }\Bbb{S}_{\nu }+2a\left( \Bbb{S}F\stackrel{.}{x}%
\right) \stackrel{.}{x}^{\mu }  \nonumber \\
&&+2a^{*}\left( F\mathcal{S}\right) ^{\mu \alpha }\stackrel{.}{x}_{\alpha }%
\stackrel{.}{x}^{2},
\end{eqnarray}
which we have to compare to the Bargmann-Michel-Telegdi equation using the
proper time
\begin{equation}
\stackrel{.}{\Bbb{S}}^{\mu }=\frac{gq}{2m}F^{\mu \nu }\Bbb{S}_{\nu }+\frac{q%
}{2m}(g-2)\left( \Bbb{S}F\stackrel{.}{x}\right) \stackrel{.}{x}^{\mu }.
\label{BMT}
\end{equation}
This procedure enable us to fix the constant $a$ under the form
\begin{equation}
a=\frac{q}{4m}(g-2).
\end{equation}

The motion equation thus becomes
\begin{equation}
\stackrel{..}{x}^{\mu }=\frac{q}{m}F^{\mu \nu }\stackrel{.}{x}_{\nu }+\frac{q%
}{2m^{2}}\mathcal{S}_{\rho \sigma }\partial ^{\mu }F^{\rho \sigma }-\frac{q}{%
4m^{2}}(g-2)\stackrel{.}{x}_{\rho }\partial ^{\mu }F^{\rho \sigma }\mathcal{S%
}_{\sigma \tau }\stackrel{.}{x}^{\tau }+....,
\end{equation}
the second is
\begin{equation}
\stackrel{.}{S}^{\mu \nu }=\frac{q}{m}\left( F^{\mu }{}_{\rho }\mathcal{S}%
^{\rho \nu }-\mathcal{S}^{\mu }{}_{\rho }F^{\rho \nu }\right) +\frac{q}{2m}%
(g-2)\left( \stackrel{.}{x}^{\mu }F^{\rho \nu }{}-\stackrel{.}{x}^{\nu
}F^{\rho \mu }\right) \stackrel{.}{x}_{\rho },
\end{equation}
and the third
\begin{equation}
\stackrel{.}{\Bbb{S}}^{\mu }=\frac{q}{2m}\left\{ \left( 2+(g-2)\stackrel{.}{x%
}^{2}\right) F^{\mu \nu }\Bbb{S}_{\nu }+(g-2)\left( \Bbb{S}F\stackrel{.}{x}%
\right) \stackrel{.}{x}^{\mu }+(g-2)^{*}\left( F\mathcal{S}\right) ^{\mu
\alpha }\stackrel{.}{x}_{\alpha }\stackrel{.}{x}^{2}\right\}.
\end{equation}
If we want to be able to compare this last result with that of the equation (%
\ref{BMT}), we must place ourselves in the context of equations expressed
according to proper time
\begin{equation}
\stackrel{.}{\Bbb{S}}^{\mu }=\frac{q}{2m}\left\{ gF^{\mu \nu }\Bbb{S}_{\nu
}+(g-2)\left( \Bbb{S}F\stackrel{.}{x}\right) \stackrel{.}{x}^{\mu
}+(g-2)^{*}\left( F\mathcal{S}\right) ^{\mu \alpha }\stackrel{.}{x}_{\alpha
}\right\} .
\end{equation}

\section{Curved space case}

The generalisation of our results to the case of a curved space is direct,
we then briefly extend the formalism below by using the results of our
preceding publications\cite{NOUS2,NOUS3} .

\subsection{Direct extention of the formalism}

To use the previous formalism in the case of a curved space we have to
introduce a general space time metric $g_{\mu \nu }(x)$ in the definition of
the new covariant Hamiltonian from the usual fundamental quadratic form $%
ds^{2}$ in the following manner 
\begin{equation}
H=\frac{1}{2}m\left( \frac{ds}{d\tau }\right) ^{2}=\frac{1}{2}mg_{\mu \nu
}(x)\stackrel{.}{x}^{\mu }\stackrel{.}{x}^{\nu }.
\end{equation}

In this curved space, where as usually the tensors are the representation of
the GL(4,$\Bbb{R}$) group, we obtain for the fundamental bra\-cket relations 
\begin{equation}
\left\{ 
\begin{array}{l}
m\left[ x^{\mu },\stackrel{.}{x}^{\nu }\right] =g^{\mu \nu }(x), \\ 
\\ 
m\left[ x_{\mu },\stackrel{.}{x}_{\nu }\right] =g_{\mu \nu }(x)+x^{\rho }%
\frac{\partial g_{\mu \rho }(x)}{\partial x^{\nu }}, \\ 
\\ 
\left[ \stackrel{.}{x}^{\mu },\stackrel{.}{x}^{\nu }\right] =\frac{q}{m^{2}}%
\mathcal{F}^{\mu \nu }(x,\stackrel{.}{x})=\frac{q}{m^{2}}F^{\mu \nu }(x)-%
\frac{1}{m}(\partial ^{\mu }g^{\rho \nu }-\partial ^{\nu }g^{\rho
\mu })\stackrel{.}{x}_{\rho }, \\ 
\\ 
\left[ \stackrel{.}{x}^{\mu },\stackrel{.}{x}_{\nu }\right] =\frac{q}{m^{2}}%
F^{\mu }{}_{\nu }(x)+\frac{1}{m}\partial _{\nu }g^{\mu \rho }\stackrel{.}{x}%
_{\rho } \\ 
\\ 
\left[ \stackrel{.}{x}_{\mu },\stackrel{.}{x}_{\nu }\right] =\frac{q}{m^{2}}%
F_{\mu \nu }(x). \\ 
\\ 
\left[ f(x,\stackrel{.}{x}),h(x,\stackrel{.}{x})\right] =\left\{ f(x,%
\stackrel{.}{x}),h(x,\stackrel{.}{x})\right\} +\frac{q}{m^{2}}\mathcal{F}%
^{\mu \nu }(x,\stackrel{.}{x})\frac{\partial f(x,\stackrel{.}{x})}{\partial 
\stackrel{.}{x}^{\mu }}\frac{\partial h(x,\stackrel{.}{x})}{\partial 
\stackrel{.}{x}^{\nu }}
\end{array}
\right.
\end{equation}
We also remark that the third equation, which defines an
electromagnetic generalized curvature, can be written as
\begin{equation}
\mathcal{F}^{\mu \nu }(x,\stackrel{.}{x})=F^{\mu \nu }(x)-\frac{m}{q}\left\{
\left( \Gamma ^{\mu \rho \upsilon }-\Gamma ^{\rho \upsilon \mu }\right) +2%
(K^{\mu }{}^{\nu \rho }+\Omega ^{\mu \rho \nu })\right\} \stackrel{.%
}{x}_{\rho }
\end{equation}
where introduce the Cartan's torsion and contorsion defined by
\begin{equation}
\left\{ 
\begin{array}{c}
\Omega ^{\mu \nu \rho }=\frac{1}{2}\left( \Gamma ^{\mu \nu \rho }-\Gamma
^{\nu \mu \rho }\right) \\ 
\\ 
K^{\mu }{}^{\nu \rho }=\Omega ^{\mu \nu \rho }+\Omega ^{\rho \mu \nu
}-\Omega ^{\nu \rho \mu }
\end{array}
\right.
\end{equation}
which are naturally null if the affine connection has its two first
indices symmetrical (case of the Christoffel symbols).

\subsection{The Motion equations}

If we want to generalize the precedent study we must introduce in the
covariant Hamiltonian the triad formalism under the simplest form
\begin{equation}
H=\frac{1}{2}mg_{\mu \nu }(x)\stackrel{.}{x}^{\mu }\stackrel{.}{x}^{\nu }-%
\frac{q}{2m}F^{ab}\mathcal{S}_{ab},
\end{equation}
the greek indices correspond to curved space and the roman indices to
Minkowski flat space, and where the infinitesimal line element is given by
\begin{equation}
ds^{2}=\eta _{ab}d\xi ^{a}d\xi ^{b}=\eta _{ab}e_{\mu }^{a}(x)e_{\nu
}^{b}(x)dx^{\mu }dx^{\nu }=g_{\mu \nu }(x)dx^{\mu }dx^{\nu }.
\end{equation}

On this level of the construction of our formalism and with the sight of the
expression of our new Hamiltonian, we continue to use the bracket between
the spinning angular momentum and the position equal to zero but it cannot
be the case for the bracket between this spinning angular momentum and the
velocity, so we suppose that there exist coefficients such as

\begin{equation}
m\left[ \mathcal{S}^{ab},\stackrel{.}{x}_{\mu }\right] =\omega _{\mu }^{ac}%
\mathcal{S}_{c}{}^{b}+\omega _{\mu }^{cb}\mathcal{S}^{a}{}_{c}{},
\label{commutation}
\end{equation}
then we have
\begin{eqnarray}
\stackrel{.}{\mathcal{S}}^{ab} &=&m\left[ \mathcal{S}^{ab},\stackrel{.}{x}%
_{\mu }\right] \stackrel{.}{x}^{\mu }-\frac{q}{2m}\left[ \mathcal{S}^{ab},%
\mathcal{S}^{cd}\right] F_{cd}  \nonumber \\
&=&\left( \omega _{\mu }^{ac}\mathcal{S}_{c}{}^{b}+\omega _{\mu }^{cb}%
\mathcal{S}{}^{a}{}_{c}\right) \stackrel{.}{x}^{\mu }-\frac{q}{m}\left( 
\mathcal{S}F-F\mathcal{S}\right) ^{ab}.  \label{temps}
\end{eqnarray}
This expression is transform in
\begin{equation}
\frac{D\mathcal{S}^{ab}}{D\tau }=\stackrel{.}{\mathcal{S}}^{ab}-\left(
\omega _{\mu }^{ac}\mathcal{S}_{c}{}^{b}+\omega _{\mu }^{cb}\mathcal{S}%
{}^{a}{}_{c}\right) \stackrel{.}{x}^{\mu }=-\frac{q}{m}\left( \mathcal{S}F-F%
\mathcal{S}\right) ^{ab}.  \label{temps S}
\end{equation}
This result is coherent with the expression found in the case of a flat
space.

The motion equation is then directly obtained by
\begin{equation}
mg_{\mu \nu }\stackrel{..}{x}^{\nu }=mg_{\mu \nu }\left[ \stackrel{.}{x}%
^{\nu },H\right] =-mg_{\mu \nu }\Gamma _{\nu \rho }^{\mu }\stackrel{.%
}{x}^{\nu }\stackrel{.}{x}^{\rho }+qF^{\mu \nu }\stackrel{.}{x}%
_{\upsilon }+\frac{q}{2}g_{\mu \nu }\left[ \stackrel{.}{x}^{\nu },F^{ab}%
\mathcal{S}_{ab}\right] ,
\end{equation}
that is to say
\begin{eqnarray}
mg_{\mu \nu }\frac{D\stackrel{.}{x}^{\nu }}{D\tau } &=&qF^{\mu \nu }%
\stackrel{.}{x}_{\upsilon }+\frac{q}{2m}\left( \partial _{\mu }F^{ab}%
\mathcal{S}_{ab}-\omega _{\mu }^{ac}F_{c}{}^{b}\mathcal{S}_{ab}-\omega _{\mu
}^{cb}F^{a}{}_{c}\mathcal{S}_{ab}\right)  \nonumber \\
&=&qF^{\mu \nu }\stackrel{.}{x}_{\upsilon }+\frac{q}{2m}D_{\mu
}F^{ab}\mathcal{S}_{ab},
\end{eqnarray}
where we observe that $\omega _{\mu }^{ab}$ are nothing other than the
components of the spin connection. The covariant derivative for a
vector in the flat metric is then given by
\begin{equation}
D_{\mu }V^{a}=\partial _{\mu }V^{a}-\omega _{\mu b}^{a}V^{b},
\end{equation}
and for the basis triads by
\begin{equation}
D_{\mu }e_{\rho }^{a}(x)=\partial _{\mu }e_{\rho }^{a}(x)V^{a}-\omega _{\mu
}^{ab}e_{\rho }^{b}(x)+\Gamma _{\mu \rho }^{\sigma }e_{\sigma }^{a}(x).
\end{equation}

We thus remark for the motion equations that the transition from a flat
space to a curved space is done only, as usually, by replacing the partial
derivative
by the covariant derivative. If we want to study the case of a space with
torsion, we must then introduce in the covariant hamiltonian a term
proportional to the electromagnetic generalised curvature $\mathcal{F}^{\mu
\nu }(x,\stackrel{.}{x})$ in the place of $F^{\mu \nu }(x)$.

\section{Link with quantum theory}

If one wants to link our classical approach to the quantum
theory it would be necessary to use a formalism with anticommuting
Grassmannian variables as Van Holten already made
\begin{equation}
\mathcal{S}^{ab}=\psi ^{a}\psi ^{b}.
\end{equation}
This Grassmannian variables satisfy the algebra relations
\begin{equation}
\left[ \psi ^{a}\psi ^{b},\psi ^{c}\psi ^{d}\right] =\eta ^{bc}\psi ^{a}\psi
^{d}-\eta ^{bd}\psi ^{a}\psi ^{c}+\eta ^{ad}\psi ^{b}\psi ^{c}-\eta
^{ac}\psi ^{b}\psi ^{d},
\end{equation}
from where we draw the commutation laws
\begin{equation}
\left\{ 
\begin{array}{c}
\left[ \psi ^{a},\psi ^{b}\right] =\eta ^{ab} \\ 
\\ 
\\ 
\left[ \psi ^{a}\psi ^{b},\psi ^{c}\right] =\left[ \psi ^{b},\psi
^{c}\right] \psi ^{a}-\left[ \psi ^{a},\psi ^{c}\right] \psi ^{b} \\ 
\\ 
\left[ \psi ^{a},\psi ^{b}\psi ^{c}\right] =\left[ \psi ^{a},\psi
^{b}\right] \psi ^{c}-\left[ \psi ^{a},\psi ^{c}\right] \psi ^{b}
\end{array}
\right.
\end{equation}
From the equations (\ref{commutation}) and (\ref{temps}) we
derive
\begin{equation}
\left\{ 
\begin{array}{c}
m\left[ \psi ^{a},\stackrel{.}{x}_{\mu }\right] =\omega _{\mu }^{ab}\psi _{b}
\\ 
\\ 
\stackrel{.}{\psi }^{a}=\omega _{\mu }^{ab}\psi _{b}\stackrel{.}{x}^{\mu }-%
\frac{q}{m}F^{ab}\psi ^{b}
\end{array}
\right.
\end{equation}
The brackets for expandable functions of $x,$ $\stackrel{.}{x},\psi $ or $%
\stackrel{.}{\psi }$ is then {\it a priori} given by
\begin{eqnarray}
\left[ f(x,\stackrel{.}{x},\psi ,\stackrel{.}{\psi }),h(x,\stackrel{.}{x}%
,\psi ,\stackrel{.}{\psi })\right]  &=&\left\{ f,h\right\} +\frac{q}{m^{2}}%
\mathcal{F}^{\mu \nu }\frac{\partial f}{\partial \stackrel{.}{x}^{\mu }}%
\frac{\partial h}{\partial \stackrel{.}{x}^{\nu }}+\eta ^{ab}\frac{\partial f%
}{\partial \psi ^{a}}\frac{\partial h}{\partial \psi ^{b}}  \nonumber \\
&&+\alpha ^{ab}\left( \frac{\partial f}{\partial \psi ^{a}}\frac{\partial h}{%
\partial \stackrel{.}{\psi }^{b}}-\frac{\partial f}{\partial \stackrel{.}{%
\psi }^{a}}\frac{\partial h}{\partial \psi ^{b}}\right)   \nonumber \\
&&+\beta ^{\mu a}\left( \frac{\partial f}{\partial x^{\mu }}\frac{\partial h%
}{\partial \psi ^{a}}-\frac{\partial f}{\partial \psi ^{a}}\frac{\partial h}{%
\partial x^{\mu }}\right)   \nonumber \\
&&+\gamma ^{\mu a}\left( \frac{\partial f}{\partial \stackrel{.}{x}^{\mu }}%
\frac{\partial h}{\partial \psi ^{a}}-\frac{\partial f}{\partial \psi ^{a}}%
\frac{\partial h}{\partial \stackrel{.}{x}^{\mu }}\right)   \nonumber \\
&&+\zeta ^{\mu a}\left( \frac{\partial f}{\partial x^{\mu }}\frac{\partial h%
}{\partial \stackrel{.}{\psi }^{a}}-\frac{\partial f}{\partial \stackrel{.}{%
\psi }^{a}}\frac{\partial h}{\partial x^{\mu }}\right)   \nonumber \\
&&+\rho ^{\mu a}\left( \frac{\partial f}{\partial \stackrel{.}{x}^{\mu }}%
\frac{\partial h}{\partial \stackrel{.}{\psi }^{a}}-\frac{\partial f}{%
\partial \stackrel{.}{\psi }^{a}}\frac{\partial h}{\partial \stackrel{.}{x}%
^{\mu }}\right) 
\end{eqnarray}
where we take $\beta ^{\mu a}=0$, because the spinning angular momentum
commute always with the position and $\gamma ^{\mu a}=-\frac{1}{m}\omega
_{\mu }^{ab}\psi _{b}$.

Now the covariant hamiltonian $H(x,\stackrel{.}{x}%
,\psi ,\stackrel{.}{\psi })$ is written 
within the framework of a ``kind of
supersymmetrical'' theory, to go further we should add a Dirac-like
term \cite{VAN HOLTEN}
according to that new variables
\begin{equation}
H(x,\stackrel{.}{x},\psi ,\stackrel{.}{\psi })=H_{0}(x,\stackrel{.}{x}%
)+H_{D}(x,\stackrel{.}{x},\psi ,\stackrel{.}{\psi }).
\end{equation}
but this would be a little apart from the matter of our simple application.

\section{Conclusion}

The goal of this work was to study the classical approach of a spinning
particle associated to a Lorentz covariant Hamiltonian. For this we have used a
four dimensional brackets structure which gives a structure of an algebra
between the position and velocity and which 
generalizes the Poisson brackets. We
have proposed
an equation of evolution for the spin tensor which generalizes in a
direct way that of Bargmann-Michel-Telegdi using the proper time,
while we have placed
ourselves in context of a time considered as the conjugate coordinate of the
covariant Hamiltonian. This kind of idea is not new, we recall that
introduced by Poincar\'{e}\cite{POINCARE2}, the proper time was imposed by
Minkowski\cite{MINKOWSKI} which showed that it is the only variable
connected with the source which is available to all observers. This
allows to propose that space and time should be unified in a single space.
Dirac\cite{DIRAC} was one of the first which raised the fact that the manner
of choosing the source proper time puts in second position the role of the
coordinate time which is the proper-time of the observer. Then many
physicists, in the goal to circumvent this problem, started to make
alternative approachs as Horwitz and Piron\cite{PIRON} in assuming a
fifth time observation parameter, historical time, connected to all
observers, or Gill, Lindesay and Zachary\cite{GILL1,GILL2} in constructing
the canonical time wich is available to all experimentaters in their frame
of reference, it is in this context that we have introduce our time
parameter.

Our formalism can be also directly extrapolated to the case of presence of a
Poincar\'{e} generalized tensor and to the curved space situation, where
the principal notions are introduced in a natural manner.

\end{document}